\documentclass[aps,prb,amsmath,amssymb,footinbib,showpacs,twocolumn,superscriptaddress]{revtex4-1}
\usepackage{amsmath}
\usepackage{amssymb}
\usepackage{amsthm}
\usepackage{setspace}
\usepackage{graphicx}
\usepackage{braket}
\usepackage{mathrsfs}
\usepackage{physics}
\usepackage{float}
\usepackage[colorlinks = true,linkcolor = red,urlcolor  = blue,citecolor = blue,anchorcolor = blue]{hyperref}
\usepackage[utf8]{inputenc}
\usepackage[english]{babel}
\usepackage{bm}
\usepackage{xcolor}

\begin{document}

\title{Curvature of gap-closing features and the extraction of Majorana nanowire parameters}

\author{Haining Pan}
\affiliation{Condensed Matter Theory Center and Joint Quantum Institute, Department of Physics, University of Maryland, College Park, Maryland 20742, USA}

\author{Jay D. Sau}
\affiliation{Condensed Matter Theory Center and Joint Quantum Institute, Department of Physics, University of Maryland, College Park, Maryland 20742, USA}

\author{Tudor D. Stanescu}
\affiliation{Condensed Matter Theory Center and Joint Quantum Institute, Department of Physics, University of Maryland, College Park, Maryland 20742, USA}
\affiliation{Department of Physics and Astronomy, West Virginia University, Morgantown, West Virginia 26506}

\author{S. Das Sarma}
\affiliation{Condensed Matter Theory Center and Joint Quantum Institute, Department of Physics, University of Maryland, College Park, Maryland 20742, USA}

\begin{abstract}

Recent tunneling conductance measurements of Majorana nanowires show a strong variation in the magnetic-field dependence of the superconducting gap among different devices. Here, we theoretically study the magnetic field dependence of the gap-closing feature and establish that the degree of convexity (or concavity) of the gap closing as a function of Zeeman field can provide critical constraints on the underlying microscopic parameters of the semiconductor-superconductor hybrid system model. Specifically, we show that the gap-closing feature is entirely concave only for strong spin-orbit coupling strength relative to the chemical potential. Additionally, the non-linearity (i.e., concavity or convexity) of the gap closing as a function of magnetic field complicates the simple assignment of a constant effective $ g $-factor to the states in the Majorana nanowire. We develop a procedure to estimate the effective $ g $-factor from recent experimental data that accounts for the nonlinear gap closing, resulting from the interplay between chemical potential and spin-orbit coupling. Thus, measurements of the magnetic field dependence of the gap closure on the trivial side of the topological quantum phase transition can provide useful information on parameters that are critical to the theoretical modeling of Majorana nanowires.
\end{abstract}
	
\date{\rm\today}
\maketitle

\section{introduction}\label{sec:intro}

The search for Majorana zero modes (MZMs) in solid state systems,\cite{nayak2008nonabelian,sarma2015majorana,alicea2012new,elliott2015colloquium,stanescu2013majorana,leijnse2012introduction,beenakker2013search,lutchyn2018majorana,aguado2017majorana,jiang2013nonabelian,sato2016majorana,sato2017topological,plugge2017majorana,karzig2017scalable,wilczek2012quantum,sau2010nonabelian} a major component of the recent quantum information upheaval in condensed matter physics,  is characterized by a perhaps perplexing dichotomy between remarkable experimental advances \cite{moor2018electric,grivnin2018concomitant,vaitiekenas2018effective,gul2018ballistic,kammhuber2016conductance,kammhuber2017conductance,das2012zerobias,deng2016majorana,albrecht2016exponential,mourik2012signatures,chen2017experimental} and the lack of unambiguous demonstration of topological superconductivity and topologically protected MZMs, as predicted within the robust framework of the  noninteracting theory.\cite{alicea2012new,stanescu2013majorana,dassarma2012splitting,oreg2010helical} Focusing on the most promising MZM platform, the semiconductor-superconductor (SM-SC) hybrid nanowire system,\cite{sau2010generic,fu2008superconducting,leijnse2012introduction,lutchyn2018majorana,jiang2013nonabelian,karzig2017scalable,lutchyn2010majorana} one notes that the overwhelming majority of experimental signatures consistent with the presence of MZM were obtained using local probes (more specifically, charge tunneling measurements \cite{sengupta2001midgap,setiawan2015conductance,reeg2017transport}); so far, these signatures are not corroborated by observations of any corresponding nonlocal feature,\cite{alicea2012new,nayak2008nonabelian,kitaev2003faulttolerant,freedman2003topological,dassarma2005topologically,bonderson2008measurementonly,fu2010electron} as predicted theoretically. It should be emphasized that non-local correlations are the hallmark of a topological system, and it is insufficient to infer about MZMs based only on local tunneling measurements. Furthermore, it has been argued that (local) signatures similar to  those generated by the presence of topological MZMs can also emerge from low-energy, non-topological Andreev bound states (ABS) in systems with smooth confinement,\cite{zazunov2011coulomb,kells2012nearzeroenergy,stanescu2014nonlocality,moore2018twoterminal,adagideli2014effects,brouwer2011topological,brouwer2011probability,sau2012experimental}  disorder, \cite{adagideli2014effects,brouwer2011topological,sau2012experimental,cole2015effects,hui2015bulk,takei2013soft,sau2013density,lutchyn2012momentum,lobos2012interplay,liu2012zerobias,bagrets2012class,pikulin2012zerovoltage,lee2012zerobias,lee2014spinresolved,chang2015hard,gul2017hard,vanweperen2015spinorbit,zhang2017ballistic} or inhomogeneous potentials.\cite{liu2017andreev,prada2012transport,gul2015high} In addition, for hybrid SM-SC structures, it is rather problematic to measure  key  parameters, such as the chemical potential, the spin-orbit coupling, and the effective $g$-factor, which control the low-energy  Majorana physics,\cite{alicea2012new,stanescu2013majorana} making it difficult to assess whether or not the experimental conditions are consistent with the emergence of topological superconductivity. For example, measured system parameters (e.g., the Land\'{e} $ g $-factor, the spin-orbit coupling, the chemical potential) in isolated semiconductor nanowires do not tell us anything about the relevant parameters for the actual SM-SC hybrid nanowire structures, where the SC substrate most certainly strongly (and in an unknown and uncontrolled manner) renormalizes the system parameters. 

To address these problems, it is essential that the experimental efforts be supplemented with theoretical studies that focus on realistic experimental conditions and aim to establish a direct relationship between key model parameters and certain robust experimental features. The recent numerical studies of electrostatic effects in SM-SC hybrid structures  based on self-consistent solutions of the Schr\"{o}dinger-Poisson problem \cite{woods2018effective,mikkelsen2018hybridization,antipov2018effects} represent  an important step in this direction. In particular, this type of work is essential for understanding the proximity effect induced by the coupling between the SM and the SC and the impact of applied gate voltages on the SM-SC coupling. In turn, these effects control the values of basic parameters, such as the induced pair potential and the Land\'{e} $g$-factor. Furthermore, three-dimensional self-consistent schemes \cite{woods2018effective} are crucial for understanding systems with inhomogeneous parameters, including the tunnel barrier regions at the ends of proximitized Majorana wires and the possible formation of quantum dots inside or at the ends of a hybrid system. An important caveat with respect to such self-consistent numerical modeling is, however, that the appropriate boundary conditions (since both Schr\"{o}dinger and Poisson equations are second-order partial differential equations, boundary conditions control the actual solutions) are simply unknown for the experimental nanowires and, therefore, extensive use of the actual experimental data as compared with theoretical simulations is essential for progress.  The current paper establishes that certain experimentally observed features in the Zeeman field driven gap closing in Majorana nanowires can be useful, when compared with appropriate theoretical simulations we develop, in providing estimates for various microscopic parameters underlying the SM-SC hybrid structures.

In this paper, we focus on establishing a direct relationship between several key effective parameters in the Majorana nanowire model, including the spin-orbit coupling and the effective $g$-factor, and a robust experimental characteristic: the gap-closing feature that characterizes the generic dependence of the low-energy differential conductance on the applied magnetic field. It has been shown \cite{huang2018metamorphosis} that below the topological quantum phase transition TQPT, i.e., in the so-called trivial SC regime, clean superconducting spin-orbit-coupled nanowires are generically  characterized by finite energy in-gap intrinsic ABSs that generate a strong gap-closing feature in the tunneling spectra. The gap-closing feature associated with these intrinsic ABSs (i-ABSs) emerging in clean systems at (generic) finite values of the chemical potential always precedes \cite{huang2018metamorphosis} the emergence of the Majorana-induced zero-bias conductance peak (ZBCP). The presence of (weak) inhomogeneities (e.g., smooth confining potentials) affects the dependence of these ABS on the applied magnetic field, but, typically, does not remove them. Consequently, the associated gap-closing feature is expected to be quite generic, as confirmed experimentally by its ubiquitous presence in the measured tunneling spectra. Here, we work under the assumption of a clean (i.e., assuming the absence of any extrinsic ABS) system and show that the shape of the gap-closing feature associated with the intrinsic ABS is determined by key effective parameters, i.e., chemical potential, spin-orbit coupling, and $g$-factor, which could be estimated by fitting the experimentally measured tunneling conductance spectra. We emphasize that the details of our current analysis are valid in the clean (i.e., homogeneous) limit without complications arising from extrinsic ABSs. Generalizations that include the effects of inhomogeneities (e.g., smooth confinement potentials) are straightforward, but highly non-universal (and, therefore, perhaps not particularly illuminating). By contrast, the clean limit results presented in the current paper are quite robust and can be used as a benchmark for the effective parameters obtained by comparing experimental data and theoretical simulations.  In particular, we show that the concavity/convexity of the gap-closing feature as a function of the increasing Zeeman field is controlled by the strength of the spin-orbit coupling.  We note that our procedure of estimating the effective parameters, which is  based on the homogeneous system assumption, is useful even when the experimental system does not actually satisfy this condition (e.g., when the physics is dominated by extrinsic ABS). In this case, the procedure will lead to inconsistencies (e.g., unphysical parameter values, discrepancies with estimates based on different procedures, unphysical dependence on control parameters such as gate potentials, etc.) that will signal the ``extrinsic'' nature of the ABS responsible for the gap-closing feature.  In turn, this type of situation should immediately call into question the nature of the ZBCP that follows the gap-closing feature, since in the presence of inhomogeneities it could be associated with either MZM or ABS coalescing toward zero energy.\cite{vuik2018reproducing} We make the pristine nanowire assumption simply because our results are then universal whereas the corresponding extrinsic ABS system will be totally determined by the details of the nanowire which would vary non-universally from sample to sample. 

The remainder of this paper is organized as follows. In Sec. II, we present the theoretical model of the SM-SC hybrid system and the numerical method used in this paper. In Sec.III, we discuss the curvature of gap-closing feature and the information it can provide about the effective parameters.  The results for the effective $g$-factor, including a discussion of field-dependent effective $g$-factor, are presented in Sec. IV. In Sec.V, we use actual experimental results to extract the effective system parameters based on a simulated annealing fitting procedure. Our conclusions are presented in Sec.VI. The Appendix provides some technical details on the subtle role of spin-orbit coupling in the theory.

\section{Model and numerical method}\label{sec:model}

To properly account for the basic effects that determine the low-energy physics of SM-SC hybrid structures under realistic, laboratory conditions, minimal models have to incorporate information regarding the nanostructure size and geometry, the coupling at the SM-SC interface, as well as disorder and external potentials, including applied gate potentials. The corresponding Hamiltonian has the following generic form 
\cite{stanescu2013majorana}
\begin{equation}\label{eq:tot}
H_{\text{tot}}=H_{\text{SM}}+H_{\text{Z}}+H_{\text{V}}+H_{\text{SC}}+H_{\text{SM-SC}},
\end{equation}
where $ H_{\text{SM}} $ is the non-interacting Hamiltonian of the semiconductor component, $ H_{\text{Z}} $ describes the applied Zeeman field, $ H_{\text{V}} $ contains contributions from disorder and gate potentials, $H_{\text{SC}} $ describes the parent superconductor, and $ H_{\text{SM-SC}} $ characterizes the SM-SC coupling. In this paper, we focus on the homogeneous, clean wire regime, hence we neglect possible contributions from  disorder and gate potentials, $H_{\text{V}}=0$. We note that, in general, non-uniform potential effects can be safely ignored if the characteristic strength of the potential inhomogeneity is small compared to the characteristic energy scale for proximity-induced superconductivity (e.g., the induced pair potential). This means, as emphasized already in the Introduction, that we are ignoring any extrinsic ABS effects in the physics of the Majorana nanowires, which is the experimental goal anyway.

Further simplifications can be made in the {weak coupling} limit, when the proximity effect due to the coupling to the parent superconductor is described  by an induced pair potential $\Delta$. In this limit, the hybrid nanowire is described by the ``standard'' minimal  Bogoliubov-de Gennes (BdG) Hamiltonian,\cite{sau2010generic,lutchyn2010majorana,oreg2010helical} $\hat{H}=\frac{1}{2}\int dx ~\hat{\Psi}^\dagger(x)H_{\text{tot}}\hat{\Psi}(x) $ with 
\begin{equation}\label{eq:H}
H_{\text{tot}} = \left( -\frac{\hbar^2}{2m^*} \partial^2_x -i \alpha \partial_x \sigma_y - \mu \right)\tau_z + V_Z\sigma_x + \Delta \tau_x.
\end{equation} 
Here, $\hat{\Psi}=\left(\hat{\psi}_{\uparrow},\hat{\psi}_{\downarrow},\hat{\psi}_{\downarrow}^\dagger,-\hat{\psi}_{\uparrow}^\dagger\right)^T $ represents a position-dependent spinor, while $\vec{\bm{\sigma}}$ and $\vec{\bm{\tau}}$ denote Pauli matrices in the spin and particle-hole space, respectively. A magnetic field applied along the wire (i.e., in the $x$ direction) produces the Zeeman term $H_{\text{Z}}=V_Z\sigma_x$, while the proximity-induced superconductivity is described (in the weak coupling limit) by term $\Delta \tau_x$. Unless stated specifically, the values of the effective parameters used in the numerical calculations are\cite{lutchyn2018majorana,gul2018ballistic,kammhuber2016conductance,kammhuber2017conductance} $m^*=0.015 m_e$ (for the effective mass) and $\Delta=0.2 $meV (for the proximity-induced superconducting gap).  In addition, the Zeeman splitting is $V_Z=\frac{1}{2}\mu_B g B$, where $B$ is the applied magnetic field, $\mu_B=5.788\times10^{-5}~$eV/T is the Bohr magneton, and $g\sim4-50$ is the effective  Land\'{e} factor (with expected values within a relative wide range. \cite{vaitiekenas2018effective,das2012zerobias,deng2016majorana,albrecht2016exponential}) The total length of the SM-SC hybrid nanowire considered in the calculations is $L\sim1-2~\mu$m.\cite{grivnin2018concomitant,vaitiekenas2018effective,moor2018electric} The theory itself is independent of \textcolor{black}{the} parameter details, but any fitting to the experimental data necessitates assumptions about some of these system parameters whereas others can be extracted by comparing theory and experiment.
	
The weak coupling limit may not be appropriate for all experimental situations, particularly those involving epitaxial {aluminum(Al)} as the parent superconductor.\cite{chang2015hard,das2012zerobias,krogstrup2015epitaxy} To go beyond the {weak coupling} approximation, we consider the proximity effect more closely, within a Green's function approach.\cite{stanescu2010proximity,sau2010robustness,sau2010nonabelian} Note that, in essence, the superconducting proximity effect is due to the electrons in the SM wire penetrating inside the parent SC. To formally capture this effect, one can integrate out the SC degrees of freedom and replace the parent SC  by a self-energy term in the effective Green's function for the wire.\cite{stanescu2010proximity,sau2010robustness,reeg2017transport,liu2017andreev,stanescu2013majorana,stanescu2017proximityinduced,stanescu2013majorana} 
Explicitly, the parent SC can be described at the mean-field level by the tight-binding BdG Hamiltonian
\begin{equation}\label{eq:sc}
H_{\text{SC}}=\sum_{i,j,\sigma}^{}\left(t_{ij}^{\text{SC}}-\mu_{\text{SC}}\delta_{ij}\right)a_{i\sigma}^\dagger a_{j\sigma}+\Delta_0\sum_i\left(a_{i\uparrow}^\dagger a_{i\downarrow}^\dagger+a_{i\downarrow}a_{i\uparrow}\right),
\end{equation}
where $i$ and $j$ are site indices, $a_{i\sigma}^\dagger$ ($a_{i\sigma}$) is the creation (annihilation) operator for an electron with spin $\sigma$ at the position $i$, $\mu_{\text{SC}}$ represents the chemical potential, and $\Delta_0$ is the parent superconducting gap. In addition, the SM-SC coupling term has the form
\begin{equation}\label{eq:sm-sc}
H_{\text{SM-SC}}=\sum_{i,j}^{}\sum_{\sigma}^{}\left[\tilde{t}_{i,j}^{\sigma}c_{i}^\dagger a_{j,\sigma}+\text{H.c}\right],
\end{equation}
where $i$ and $j$ label lattice sites at the SM-SC interface (inside the SM and the SC, respectively) and $\tilde{t}_{i,j}^{\sigma}=\tilde{t}$  are hopping matrix elements between nearest-neighbor interface sites. Integrating out the superconductor degrees of freedom results in a surface self-energy contribution to the semiconductor Green{'s} function,
\begin{equation}\label{eq:se}
\Sigma_{i,i}(\omega)=-\abs{\tilde{t}}^2v_F\left[\frac{\omega+\Delta_0\tau_x}{\sqrt{\Delta_0^2-\omega^2}}+\xi\tau_z\right],
\end{equation}
where $v_F$ is the surface density of states of the superconductor at the Fermi energy and $ \xi $ is a proximity-induced shift of the chemical potential. In the numerical calculations we take $ \xi=0 $ and include this contribution in the effective chemical potential. The corresponding frequency-dependent total ``Hamiltonian'' that generalizes the expression in Eq.\eqref{eq:H} is
\begin{equation}\label{eq:SE}
H_{\text{SE}}=\left( -\frac{\hbar^2}{2m^*} \partial^2_x -i \alpha \partial_x \sigma_y - \mu \right)\tau_z + V_Z\sigma_x -\gamma\frac{\omega+\Delta_0\tau_x}{\sqrt{\Delta_0^2-\omega^2}},
\end{equation}
where the effective SM-SC coupling is $\gamma=\abs{\tilde{t}}^2v_F$. Note that in the weak coupling regime ($\gamma\ll \Delta_0$) the pairing term becomes $\Delta\tau_x$, with an induced gap $\Delta=\frac{\gamma\Delta_0}{\gamma+\Delta_0}\sim \gamma$, while the dynamical correction (i.e., the term proportional to $\omega$) can be neglected. 
\cite{stanescu2017proximityinduced,stanescu2013majorana,lutchyn2011search,stanescu2011majorana} \textcolor{black}{Although the two models provide similar results qualitatively, there are significant quantitative differences. We emphasize that value of critical $ V_Z $ necessary for closing the gap at the TQPT is much higher in the self-energy model in the strong-coupling regime.} Within this approximation, Eq. \eqref{eq:SE} reduces to the effective Hamiltonian in Eq.\eqref{eq:H}. Below, we  investigate the intermediate/strong coupling regime with $\gamma$ comparable to the superconducting gap for {Al}, \cite{moor2018electric} $\Delta_0=0.34~$meV. 
	
We calculate numerically the low-energy spectra of tight-binding models obtained by discretizing the continuous Hamiltonians in Eqs.\eqref{eq:H} and \eqref{eq:SE}.\cite{dassarma2016how} The tight-binding Hamiltonians are diagonalized for different values of the Zeeman splitting $ V_Z $ using the Arnoldi iteration technique\cite{arnoldi1951principle} for sparse matrices, which is an efficient way of obtaining the lowest energy in particle channel  and, thus, the gap closing line. Note that the self-energy ``Hamiltonian'' Eq. \eqref{eq:SE} is $ \omega $-dependent, which means that a straightforward diagonalization is not possible and an iterative method has to be considered instead. 

We conclude this section with a summary of the main approximations used in the construction of the tight-binding model. First, we work in the homogeneous, clean wire limit, so we neglect all possible contributions from disorder and gate-induced potentials. Second, the effects of many-body interactions are not considered explicitly, but are incorporated into the effective model parameters. Third, we work in the single-band approximation, which holds as long as the occupancy of the wire is low and there are no strong potential inhomogeneities. Note that for nanowires having many occupied bands (of the order 10-30) and strong inhomogeneities, the inter band couplings become significant and the single-band approximation fails.\cite{lutchyn2011search,stanescu2011majorana} But, such a multi band situation is manifestly non-universal in a complicated manner. Finally, the SM-SC coupling is treated both in the weak coupling (static) approximation and, more accurately, within a self-energy approach that captures the proximity-induced pairing as well as the proximity-induced low-energy renormalization. We note that the most significant impact on the gap-closing feature would result from breaking the homogeneous wire assumption, while relaxing the other approximations is expected to generate relatively minor quantitative changes. Therefore, any significant inconsistency of the results obtained using the procedure described in this paper should be naturally interpreted as an indication that the low-energy physics of the hybrid system is most likely controlled by inhomogeneities (e.g., disorder, smooth confining potentials, unwanted quantum dots, etc.). In such a situation (i.e., when our simulations lead to inconsistencies as compared with the experimental data), the unfortunate conclusion has to be that the nanowire physics is dominated by non-topological ABS rather than topological MZM.

\section{The curvature of the gap-closing feature}\label{sec:curvature}

In this section, we relate the measured convexity or concavity of the gap-closing feature to potential constraints on the parameters of the Majorana nanowire model. For ideal wires (i.e., without end quantum dots), the visible conductance peak that produces the gap-closing feature is associated with i-ABSs, which are generically present at finite chemical potential.\cite{huang2018metamorphosis} Therefore, the degree of concavity of the gap closure can be determined from the dependence of the lowest energy state on the applied Zeeman field in the topologically trivial regime, i.e., $ E(V_Z) $ for $0 < V_Z < V_{Zc}$, where 
$E(V_Z)>0$ is the (positive) energy of the i-ABS. The curvature of the gap-closing feature is determined by the  second derivative of $E(V_Z)$, with $\frac{\partial^2 E}{\partial V_Z^2}\ge 0~(\le 0)$ representing a convex (concave) feature. Such convexity or concavity is routinely present in the experimental gap-closing features of Majorana nanowires, but is rarely commented upon in the literature where the focus is almost always on the ZBCP beyond the full closing of the gap.

\begin{figure*}[t]
\includegraphics[width=17.8cm]{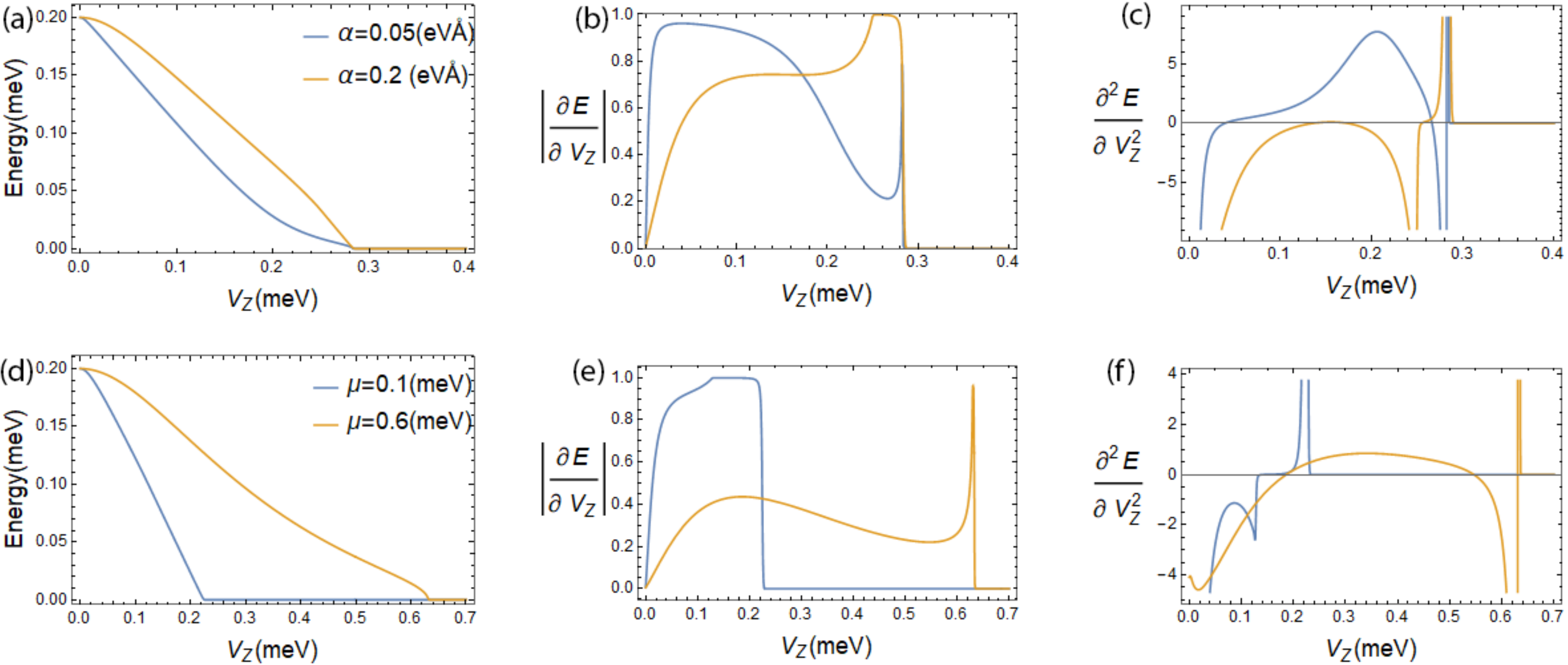}
\caption{(Color online) (a) Energy of the lowest-lying state as a function of  Zeeman field for a system with $ \alpha=0.05~$ eV\AA$~$ (blue) and $\alpha=0.2~$eV\AA$~$(orange) within the {weak coupling} model defined by Eq. (\ref{eq:H}) with $\Delta=0.2~$meV, $\mu=0.2~$meV, and wire length $L=30~\mu$m. The orange line is mostly concave while the blue line is mostly convex. (b) Absolute value of the first derivative of the lowest energy modes shown in (a).  (c) Second derivative of the lowest energy modes shown in panel (a).  The blue line is mostly positive, indicating a predominantly convex gap-closing feature, while the orange line corresponds to a concave gap-closing feature.
(d) Lowest energy as a function of Zeeman field for a {weak coupling} model with $\Delta=0.2~$meV, $\alpha=0.2~$eV\AA, $L=30~\mu$m, and two different values of the chemical potential: $\mu=0.1~$meV (blue) and $\mu=0.6~$meV (orange). The orange line is mostly convex while the blue line is almost linear.  (e) First derivative of the lowest energy modes shown in (d).  (f) Second derivative of the lowest energy modes shown in panel (d).} \label{fig:1}
\end{figure*}

We investigate the dependence of the curvature of the gap-closing feature on the effective parameters starting with the simple (weak coupling) model defined by Eq. \ref{eq:H}.
The results of the numerical calculation, including the dependence of the lowest energy on the applied Zeeman field and the first two derivatives of $E(V_Z)$, are shown in  Fig.\ref{fig:1}. Notice that, in general, the energy of the i-ABS does not have a linear dependence on the Zeeman field, i.e., the curvature of $E(V_Z)$ is nonzero, which complicates a simple definition of the $ g $-factor. Moreover,  a system with relatively weak spin-orbit coupling, $\alpha=0.05~$eV\AA$~$[blue lines in Fig.\ref{fig:1}(a)-(c)], has a preponderantly convex gap-closing feature, i.e., $\frac{\partial^2 E}{\partial V_Z^2}\ge 0$ for values of the Zeeman field $V_Z \in \mathbb{V}$ that are not too close to $V_Z=0$ or $V_Z=V_{Zc}$, {where $ \mathbb{V} $ is the interval of $ V_Z $ that satisfies $\frac{\partial^2 E}{\partial V_Z^2}\ge 0$}. By contrast, in the presence of a stronger spin-orbit coupling ($\alpha=0.2~$eV\AA, orange lines) the gap-closing feature is, basically, concave. Note that the small convex region in the vicinity of {TQPT} at $V_{Zc}=\sqrt{\mu^2+\Delta^2}$ is a finite size effect,\cite{cole2017ising,mishmash2016approaching} which is always present because of the finite nanowire length. 
In addition to the dependence on the  spin-orbit coupling strength, the curvature of the gap-closing feature depends on the chemical potential $ \mu $, as shown in the lower panels of  Fig. \ref{fig:1}. In general, we find that for a given value of the spin-orbit coupling the  gap-closing feature becomes less concave and, eventually, partially convex, with increasing chemical potential. This trend is illustrated by the comparison between the blue lines  
($ \mu=0.1~$meV) and the orange lines ($ \mu=0.6~$meV) in the lower panels of  Fig.\ref{fig:1}. \textcolor{black}{Note that we are using an extremely long nanowire($ 30~\mu m $) for the reason that we only intend to reveal the gap-closing feature dependence of chemical potential $ \mu $ and spin-orbit coupling $ \alpha $. Using a shorter length will in general increase the amplitude of Majorana oscillations beyond the TQPT significantly, the amplitude is also determined by chemical potential $ \mu $ and spin-orbit coupling $ \alpha $.\cite{fleckenstein2018decaying} In Sec.\ref{sec:experiment}, where we used several latest experiment data, \cite{moor2018electric,grivnin2018concomitant,vaitiekenas2018effective} we do not see a noticeable oscillation in our result. Our main motivation for using long wire lengths is to focus on the gap closing physics below TQPT without any complications arising from finite size effects or Majorana oscillations.}


\begin{figure}[hbtp]
	\includegraphics[width=8.6cm]{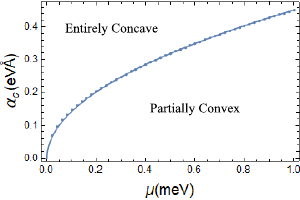}
	\caption{(Color online) Dependence of the ``critical'' spin-orbit coupling $\alpha_c $ on the chemical potential within a {weak coupling} model with $\Delta=0.2~$meV. Spin-orbit coupling values above $\alpha_c$ result in a completely concave gap-closing feature, while $\alpha < \alpha_c$ corresponds to partially convex features. Note that the analytical result given by Eq. (\ref{eq:alphac}) ({solid} line) is in excellent agreement with the numerical calculation ({dots}). 
		}
	\label{fig:2}
\end{figure}

\begin{figure*}[t]
\includegraphics[width=17.8cm]{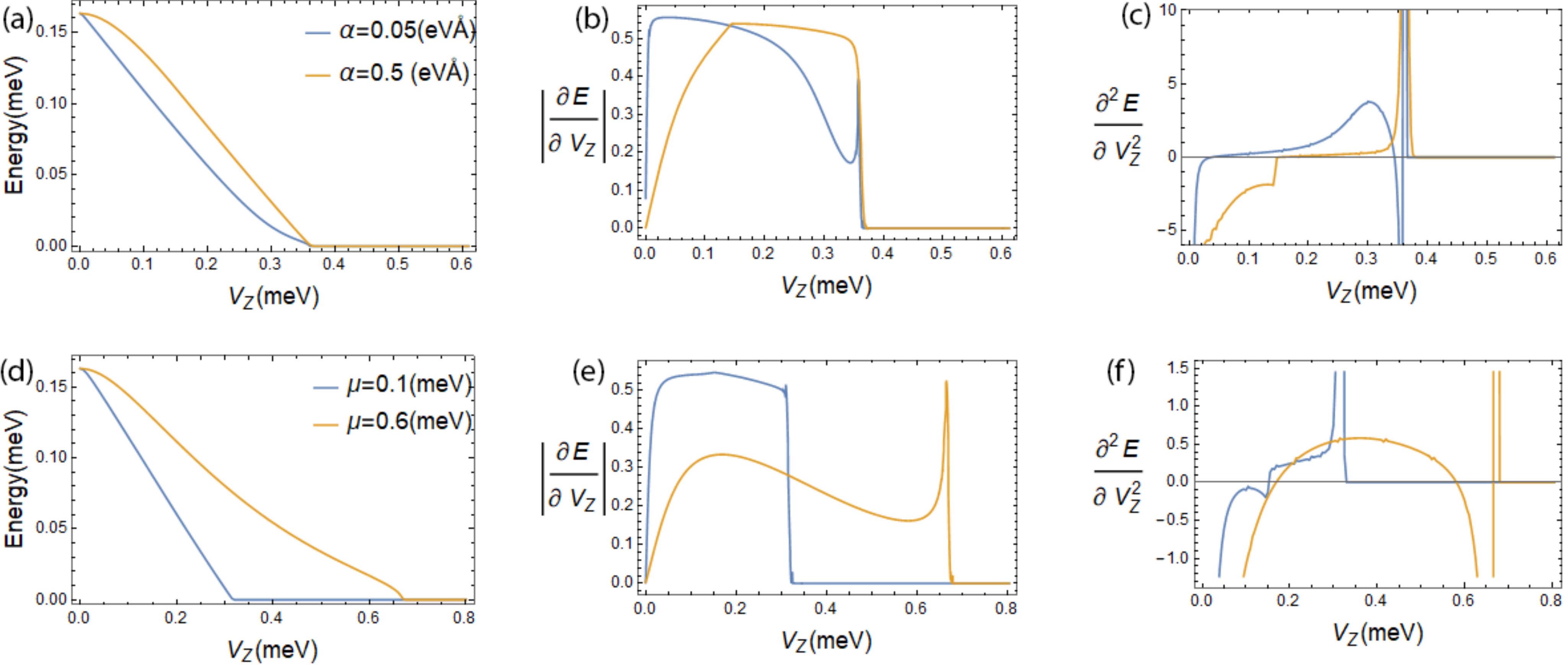}
\caption{(Color online) (a) Energy of the lowest-lying state as a function of  Zeeman field for the self-energy model defined by Eq. (\ref{eq:SE}) with $\Delta_0=0.3~$meV, $\gamma=0.3~$meV, $\mu=0.2~$meV, $L=30~\mu$m, and two values of the spin-orbit coupling: $\alpha=0.05~$eV\AA$~$ (blue) and $\alpha=0.5~$eV\AA$~$ (orange).  (b) Absolute value of the first derivative of the lowest energy modes shown in (a).  (c) Second derivative of the lowest energy modes shown in panel (a).  The blue line is mostly positive, indicating a predominantly convex gap-closing feature, while the orange line corresponds to partially concave gap-closing feature that is nearly linear over a significant $V_Z$ range.
(d) Lowest energy as a function of Zeeman field for a self-energy model with $\Delta_0=0.3~$meV, $\gamma=0.3~$meV, $\alpha=0.2~$eV\AA, $L=30~\mu$m, and two values of the chemical potential: $\mu=0.1~$meV (blue) and $\mu=0.6~$meV (orange).  The orange line is mostly convex while the blue line is approximately linear over a significant $V_Z$ range .  (e) First derivative of the lowest energy modes shown in (d).  (f) Second derivative of the lowest energy modes shown in panel (d).}
\label{fig:3}
\end{figure*}

To determine the ``critical'' value $\alpha_c $ of the spin-orbit coupling associated with the gap-closing feature becoming entirely concave, we introduce the ancillary function\cite{huang2018metamorphosis} {$ F(E) $ (see Appendix A)}. In particular, for $\alpha=0$ we have $E(V_Z)=(\Delta-V_Z)\theta(\Delta-V_{Z}) $,  where $ \theta(x) $ is Heaviside step function, and the gap-closing feature is linear $V_Z \in [0,V_{Zc}]$. However, for infinitesimal values of the spin-orbit coupling  $\alpha\neq 0$ the gap-closing feature becomes convex over the whole range $0 < V_Z < V_{Zc}$. Upon further increasing $\alpha$, the gap-closing feature remains convex only within a certain (shrinking) range $V_Z \in \mathbb{V}\subset (0,V_{Zt})$ and, eventually, becomes completely concave at $\alpha_c$. Here, $V_{Zt} < V_{Zc}$ is the crossover point from the i-ABS regime to the bulk state regime.\cite{huang2018metamorphosis} {The resultant value of critical spin-orbit coupling $ \alpha_c $ is}  (see Appendix A) 
\begin{equation}\label{eq:alphac}
\alpha_c=\beta\sqrt{\frac{h^2\mu}{m^*}}  ,
\end{equation} 
where the dimensionless coefficient $ \beta $ can be determined numerically. 

The dependence of the critical spin-orbit coupling on the chemical potential for a system with $\Delta=0.2~$meV is shown in Fig. \ref{fig:2}. Note that the analytical result (solid line) given by Eq. (\ref{eq:alphac}) is in excellent agreement with the fully numerical result (dots). The monotonic dependence  $\alpha_c\propto\sqrt{\mu}$ provides an explanation for the trend discussed in relation to Fig. \ref{fig:1}, i.e., that for a given spin-orbit coupling the  gap-closing feature becomes less concave and, eventually, partially convex, with increasing chemical potential. Also, for a given value of the chemical potential, the gap-closing feature is completely concave if $\alpha > \alpha_c(\mu)$ and becomes more concave with increasing $\alpha$. On the other hand, $\alpha < \alpha_c(\mu)$ the  gap-closing feature is partially convex (and becomes more convex upon reducing the spin-orbit coupling). Thus, convexity/concavity of gap-closing features contains rich implicit information about the underlying spin-orbit coupling, $ g $-factor (which converts the applied magnetic field to a Zeeman energy splitting), and chemical potential.

Next, we generalize our investigation of the curvature of  the gap-closing feature by fully incorporating the proximity effect  within a self-energy approach (see Sec. \ref{sec:model}). The numerical results for a system with intermediate SM-SC coupling ($\gamma = \Delta_0=0.3~$meV) are shown in Fig.\ref{fig:3}. Note that the critical Zeeman field $V_{Zc} $ associated with the TQPT  is larger than the corresponding field in the weak coupling regime, as shown in Fig.\ref{fig:3}(a). More specifically, it is controlled by the effective SM-SC coupling, rather than the induced gap, and we have 
$V_{Zc}= \sqrt{\mu^2+\gamma^2}$ (instead  $V_{Zc}= \sqrt{\mu^2+\Delta^2}$), with $\gamma >\Delta$. While there are quantitative differences between the weak coupling results shown in Fig.\ref{fig:1} and the intermediate coupling behavior illustrated in Fig.\ref{fig:3}, one notices that the general trends are not affected by the coupling strength. In particular, the low values of the chemical potential and large spin-orbit coupling favor the emergence of concave gap-closing features, while increasing $\mu$ and decreasing $\alpha$ leads to convex features. 

We conclude that the curvature of the gap-closing feature is strongly dependent on two key system parameters: the chemical potential $\mu$ and the spin-orbit coupling $\alpha$. In addition to being a qualitative indicator of the effective parameter regime --- with (robust) concave features signaling large spin-orbit coupling and predominantly convex features being associated with weak spin-orbit coupling --- the parameter-dependent curvature  of the gap-closing feature contains quantitative information that can be extracted by appropriately fitting the theory to the experimentally measured magnetic-field-dependent differential conductance. In addition, we predict that the curvature of the gap-closing feature can be modified by changing the chemical potential in a single sample(e.g., using gate potentials). Specifically, we predict that increasing the chemical potential reduces the concavity (or enhances the convexity) of the gap-closing feature.
Note, however, that in practice it may be difficult to vary $\mu$ while maintaining a constant spin-orbit coupling (since, in principle, applied gate potentials control both $\mu$ and $\alpha$). Thus, any fitting of the theory to experiment must necessarily involve multivariable regression analysis.

\section{The effective $g$-factor}\label{sec:g}

In this section, we use our understanding of the Zeeman energy dependence of the gap-closing feature to clarify the interpretation of effective $ g $-factor used in the literature. \cite{vaitiekenas2018effective,moor2018electric,antipov2018effects,mikkelsen2018hybridization} First, we note that 
the effective $g$-factor is typically understood as a dimensionless factor entering the relation between the Zeeman splitting and the applied magnetic field, $V_Z = \frac{1}{2}\mu_B g B$. However, we emphasize that even for a bare SM wire, $ g $ represents an {\textit{effective}} parameter that, in principle, is band dependent (i.e., takes different values for different confinement-induced sub-bands) and may incorporate nonlinear effects (i.e., may have some dependence on the applied magnetic field). In addition, this ``bare'' effective parameter could be further renormalized as a result of electrostatic effects and the proximity-coupling of the wire to the parent superconductor.\cite{stanescu2010proximity} A given theoretical model may implicitly incorporate some of these effects, while others are treated explicitly, e.g., the model in Eq. (\ref{eq:H}) assumes that the renormalization due to proximity-effect is weak (and is already included in the effective model parameters), while the model in Eq. (\ref{eq:SE}) addresses this effect explicitly. In principle, a certain level of modeling  can be considered as appropriate if the effective parameters of the model can be considered as being (approximately) constant over the relevant range of control parameters. If the SM-SC coupling is strong, for example, the proximity effect has to be treated explicitly, even though the low-energy physics of the hybrid system could be obtained using a weak coupling model with field-dependent parameters. The correct level of modeling appropriate for systems studied in the laboratory can only be determined through a systematic and detailed comparison between theory and experiment. 

In the light of the preceding discussion, it is clear that one should be careful when using experimental data to extract the values of effective parameters. \cite{bommer2018spinorbit,moor2018electric,vaitiekenas2018effective} Generally speaking, obtaining a dependence on the control parameters (e.g., applied magnetic field) is the first (and most straightforward) indicator that the level of modeling used in the fitting process is not sufficient for describing the system. For example, based on a weak coupling model with $\mu=0$, the lowest energy state in the topologically trivial regime drops linearly to zero. Consequently, based on this model, the effective $g$-factor could be obtained from the gap-closing feature as  $|g|=|\tilde{g}|$, where $|\tilde{g}|= \abs{\frac{2}{\mu_B}\frac{\partial E}{\partial B}}$ is the slope function characterizing the lowest energy mode (in the topologically trivial phase). Of course, if one observes any  nonlinearity (i.e., a convex or concave gap-closing feature with nonzero curvature), one should conclude that the model is not appropriate (i.e., $\mu\neq 0$,  or the SM-SC coupling is strong, or some other effect/combination of different effects  should be considered explicitly). Note that defining the effective $g$-factor as the average slope $|\overline{{g}}|=\frac{2}{\mu_B}\abs{\overline{\frac{\partial E}{\partial B}}}$, with $ \overline{\frac{\partial E}{\partial B}}$ being the average slope of the energy as a function of the applied magnetic field  (typically taken over some range of approximate linearity), assumes implicitly that any effect not explicitly included in the {weak coupling} $\mu=0$ model is negligible. Such an assumption is unwarranted and should not be made uncritically. This type of assumption has to be checked by systematically refining the modeling (i.e., including finite chemical potential, proximity-induced effects, electrostatic effects, multi-band physics, etc.) and comparing the results with experiment. We believe that most of the existing experimental $ g $-factor estimates in Majorana nanowires are suspect because of the complications we discuss above.

In the models used in this paper the $g$-factor is not a key parameter, but rather an auxiliary parameter used for converting the Zeeman field $V_Z$ (a key control parameter in the effective models) into an experimentally measurable quantity, the magnetic field $B$. Consequently, the (effective) $g$-factor is fully meaningful as long as it is a constant (i.e., $ B $-independent).  If nonlinearities of the gap-closing feature similar to those   shown in Figs. \ref{fig:1}(b) and \ref{fig:1}(e) (based on a {weak coupling} model with finite chemical potential) and Figs. \ref{fig:3}(b) and \ref{fig:3}(e) (based on an intermediate-coupling scenario) are observed experimentally, defining an effective $g$-factor as $|\tilde{g}|=\abs{\frac{2}{\mu_B}\frac{\partial E}{\partial B}}$ is meaningless, since this definition assumes implicitly that the system can be sufficiently well described by a weak coupling model with $\mu=0$ and constant (i.e., $ B $-independent) $g$-factor. Our work demonstrates that in systems with finite chemical potential (and, possibly, in the presence of strong SM-SC coupling) the gap-closing feature --- which is the feature most directly related to the effective $g$-factor --- has non-zero curvature, which is manifestly inconsistent with the $\mu=0$, weak coupling assumption of a constant $ g $-factor.  The difference between the effective $g$-factor $g$ and the quantity $\tilde{g}$ is discussed explicitly in the next section based on fits of some recent experimental data.\cite{moor2018electric,vaitiekenas2018effective,grivnin2018concomitant}

\begin{figure}[t]
	\includegraphics[width=8.6cm]{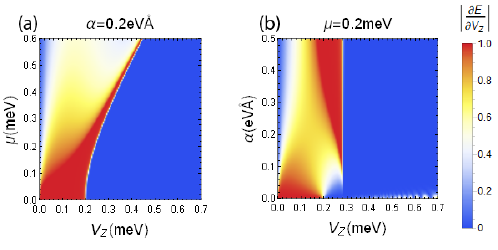}
	\caption{(Color online) (a) Absolute value of the first derivative (with respect to $ V_Z$) of the lowest energy mode as a function of Zeeman field and chemical potential for a {weak coupling} model with  $ \Delta=0.2~$meV and $ \alpha=0.2~$eV\AA. Red indicates a slope equal to 1, while blue corresponds to a vanishing slope. (b) Absolute value of the first derivative of the lowest energy mode as a function of Zeeman field and spin-orbit coupling for a {weak coupling} model with $ \Delta=0.2~$meV and $ \mu=0.2~$meV.}   
	\label{fig:4}
\end{figure}

To illustrate the dependence of the slope function $\tilde{g}$ on the magnetic field, we calculate the quantity $\abs{\frac{\partial E}{\partial V_Z}} = \abs{\frac{\tilde{g}}{g}}$ as function of the Zeeman field $V_Z$  and chemical potential $\mu$ (for a fixed value of the spin-orbit coupling $\alpha$), as well as function of $V_Z$ and $\alpha$ (for fixed $\mu$) using the {weak coupling} model given by Eq. (\ref{eq:H}). Here, the ``true'' $g$-factor $g$ is a constant that can only be determined using a more detailed model of the hybrid structure, or by comparison with experiment.
The results are shown in Figs. \ref{fig:4} (a) and 4 (b), respectively. We find that, generically,  $\tilde{g}$ is a field-dependent quantity, except for the special
 case $\mu=0$, when $\tilde{g}\equiv\overline{g}=g$.  
 For any non-zero value of the chemical potential, $\tilde{g}(V_Z)$ increases  from $\tilde{g}=0$ at $ V_Z=0 $ to $\tilde{g}=g$ at $ V_Z=V_{Zc} $. Note
  that the average slope (and, consequently, $\overline{g}$) decreases as $ \mu $ increases, which means that the discrepancy between $ g $ and $ \overline{g} $ becomes larger with increasing $\mu$. This is particularly significant if we keep in mind that the i-ABS-induced gap-closing feature is strong at large values of the chemical potential (compared to the induced gap), while it disappears at $\mu=0$. 
Finally,  if we fix $ \mu $ and investigate the dependence on the spin-orbit coupling $\alpha$, we find that $\tilde{g}$ is strongly field-dependent for any spin-orbit coupling strength. We conclude that defining the effective $g$-factor as the quantity $\tilde{g}$ (or $\overline{g}$) is meaningful only in a special parameter regime and should be generally avoided. Our theory presented in this section clearly indicates a pathway for how this problem should be approached in specific experimental situations.

\begin{figure*}[t]
\includegraphics[width=0.9\textwidth]{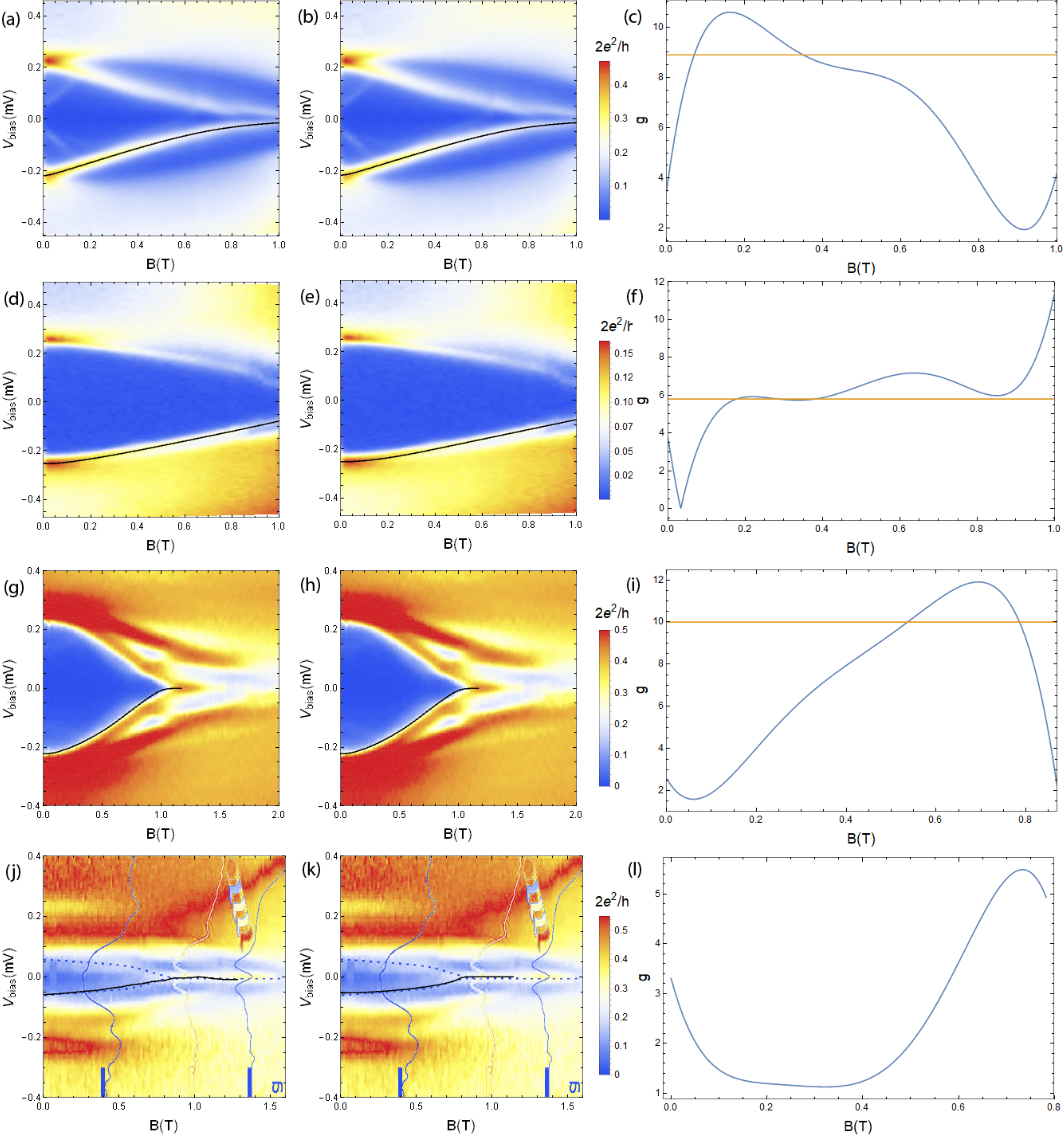}
\caption{(Color online) Fitting of gap closing signatures observed experimentally in low-energy differential conductance data (color maps) using the self-energy model given by Eq. (\ref{eq:SE}) [left panels (a), (d), (g) (j)] and the weak coupling model defined by Eq. (\ref{eq:H}) [middle panels (b), (e), (h), (k)]. The solid black lines indicates the negative energy branch of the i-ABS obtained using the fitting parameters. The right panels, (c), (f), (i), (l) show the slope function $\tilde{g}(V_Z)$ of the corresponding gap-closing feature (blue lines), as well as the average slope parameter $\overline{g}$ extracted from the (approximately) linear segments of the gap-closing features (orange). 
(a) Data from Ref. \onlinecite{moor2018electric} fitted using the self-energy model. Fitting parameters: $ \mu^\bullet=0.96~$meV, $\gamma^\bullet=0.557~$meV, $ \alpha^\bullet=0.089~$eV\AA, and $g^\bullet=25.9$, with err$=0.38$.
(b) Same data as in panel (a) fitted using the {weak coupling} model.  Fitting parameters: $\mu^\circ=0.38~$meV, $\Delta^\circ=0.22~$meV, $\alpha^\circ=0.054~$eV\AA, $g^\circ=10.4$,  with err$=0.41$. (c) Slope function $\tilde{g}(V_Z)$ for the experimental data in (a) and (b) (blue line).  The orange line corresponds to the reported\cite{moor2018electric}  linear fit,  $\overline{g}= 8.9\pm0.1$.
(d) Data from Ref. \onlinecite{moor2018electric} fitted using the self-energy model. Fitting parameters: $\mu^\bullet=0.74~$meV, $\gamma^\bullet=0.88~$meV, $ \alpha^\bullet=0.49~$eV\AA, $g^\bullet=29.5$, with err$=0.5$. 
(e) Same data as in panel (d)  fitted using the {weak coupling} model.  Fitting parameters: $\mu^\circ=0.12~ $meV, $\Delta^\circ=0.25~$meV, $\alpha^\circ=0.41~$eV\AA, $ g^\circ=6.9$, with and err$=0.67$. (f) Slope function (blue) and reported average slope\cite{moor2018electric}  $\overline{g}=5.8\pm0.2 $ (orange).
(g) Data from Ref. \onlinecite{vaitiekenas2018effective} fitted using the self-energy model. Fitting parameters: $\mu^\bullet=0.26~$meV, $\gamma^\bullet=0.25~$meV, $ \alpha^\bullet=0.65~$eV\AA, $g^\bullet=29.5$, with err$=4.6$. 
(h) Same data as in panel (g)  fitted using the {weak coupling} model.  Fitting parameters: $\mu^\circ=1.08~$meV, $\Delta^\circ=0.17~$meV, $\alpha^\circ=0.11~$eV\AA, $g^\circ=17.8$, with err$=6.18$. (i) Slope function (blue) and reported average slope\cite{vaitiekenas2018effective} $\overline{g}=10$ (orange). 
(j) Data from Ref. \onlinecite{grivnin2018concomitant}. fitted using the self-energy model. Fitting parameters:  $ \mu^\bullet=0.27~$meV, $\gamma^\bullet=0.07~$meV, $ \alpha^\bullet=0.23~$eV\AA, $ g^\bullet=10.8$, with err$=7.964$. 
(k)  Same data as in panel (j)  fitted using the {weak coupling} model.  Fitting parameters: $\mu=0.21^\circ$meV, $\Delta^\circ=0.052~$meV, $\alpha^\circ=0.51~$eV\AA, $ g^\circ=9.1$, with err$=6.1$. (l) Slope function $\tilde{g}(V_Z)$ for the data in (j) and (k).}
\label{fig:5}
\end{figure*}

\section{Fitting experimentally measured gap-closing features}\label{sec:experiment}

In this section we apply a fitting scheme based on the theoretical models described  in Sec. \ref{sec:model},  using simulated annealing to extract key parameter values from recently reported experimental results.\cite{moor2018electric,vaitiekenas2018effective,grivnin2018concomitant}
First, we collect the training data by extracting from experimental sample data, i.e, the conductance plot, by describing the position of the i-ABS-induced peak (i.e., the gap-closing feature) through a set of points of the form $ \left\{B(E_i),E_i\right\} $, where $ E_i\in\mathbb{E} $ is the energy of the lowest-lying ``visible'' mode at a magnetic field $ B(E_i) $ and  $ \mathbb{E}$ is the ordered set of energies defining the training data, from the largest to the smallest. We typically collect around 100 field-energy pairs from each experimental conductance plot. Next, we introduce the loss function
\begin{equation}\label{key}
\text{err}(\mu,\Delta,\alpha)=\sum_{i=1}^{N} \left(\frac{V_Z^{\left(\text{th}\right)}(E_i)}{V_Z^{\left(\text{th}\right)}(E_{N})}-\frac{B^{\left(\text{exp}\right)}(E_i)}{B^{\left(\text{exp}\right)}(E_{N})}\right)^2,
\end{equation}
where $E_N\neq0$ is the lowest (non-vanishing) energy from the set $ \mathbb{E}$,  (exp)  refers to the experimental data, and (th) are to quantities generated using the theoretical model. Note that  $V_Z^{\left(\text{th}\right)}(E_i)$ depends on the input parameters $ (\mu,\Delta,\alpha) $, if we use the weak coupling model from Eq. (\ref{eq:H}), or $ (\mu,\gamma,\alpha) $ in the case of the self-energy model given by Eq.  (\ref{eq:SE}).
The main reason for choosing this kind of loss function is that we do not have the $g$-factor as an additional fitting parameter, but simply assume that  $ \frac{B}{V_Z} $ is constant (i.e., $ B $-independent). The fitting parameters are determined by minimizing the the loss function: $\text{err}(\mu^\circ,\Delta^\circ,\alpha^\circ) = {\rm Min}[\text{err}(\mu,\Delta,\alpha)]$, for the {weak coupling} model, and  $\text{err}(\mu^\bullet,\gamma^\bullet,\alpha^\bullet) = {\rm Min}[\text{err}(\mu,\gamma,\alpha)]$, for the self-energy model. Finally, by estimating the experimental value of the critical field $B_c$ corresponding to the TQPT and relating it to the critical Zeeman splitting $V_{Zc}$, we obtain the fitted values of the effective $g$-factor, $g^\circ$ and $g^\bullet$, respectively. 
	
We implement the fitting procedure using the two models described in  Sec. \ref{sec:model}: (i) the {weak coupling} model defined by Eq.\eqref{eq:H} and   (ii) the self-energy nanowire model Eq. (\ref{eq:SE}). The corresponding results are shown in Fig. \ref{fig:5}(b), \ref{fig:5}(e), \ref{fig:5}(h) and \ref{fig:5}(k) and Fig. \ref{fig:5}(a), \ref{fig:5}(d), \ref{fig:5}(g) and \ref{fig:5}(j), respectively, with the fitting parameters provided in the figure caption. Several observations are warranted. First, our analysis reveals a significant discrepancy between the values of the effective $g$-factor obtained using the fitting procedure (i.e., $g^\circ$ or $g^\bullet$) and the average slope $\overline{g}$ extracted directly from the experimental data. In addition, the slope function $\tilde{g}(B)$ exhibits large variations over the relevant range of magnetic fields. For example, the gap-closing feature shown in Figs. \ref{fig:5}(a) and \ref{fig:5}(b) is characterized by a slope function [blue line in Fig. \ref{fig:5} (c)] that varies between $\tilde{g}\approx 2$ and $\tilde{g}\approx 10.5$ and, using the low-field region ($B < 0.5~$T) for the linearization procedure, one can extract\cite{moor2018electric} an average slope $\overline{g} = 8.9\pm0.1$. By contrast, the weak coupling fitting gives $g^\circ = 10$, while the fit based on the self-energy model gives $g^\bullet = 26$.  Note that, while $g^\circ$ is comparable to  $\overline{g}$, the fit based on the self-energy model (i.e.,  $g^\bullet$) is significantly different. In general, both $g^\circ$  and $g^\bullet$ can differ significantly from the average slope  $\overline{g}$, as illustrated by the results shown in Figs.\ref{fig:5} (g)-\ref{fig:5}(i).  This demonstrates that, in general, the slope of the gap-closing feature cannot be directly related to the effective $g$-factor (as is often done by the experimentalists) and should not be used as an estimate for this parameter. Our second observation concerns the (significant) difference between $g^\circ$  and $g^\bullet$. To understand this difference, one has to keep in mind that both  $g^\circ$  and $g^\bullet$ are {\textit{effective} } parameters, but $g^\circ$ includes the proximity-induced renormalization, while $g^\bullet$ does not (since the self-energy model addresses this effect explicitly). Consequently, the comparison between  $g^\bullet$ and  $g^\circ$ allows us to estimate the strength of the effective SM-SC coupling strength: Comparable values imply weak coupling, while a large discrepancy signals a strong-coupling regime. We note that the data shown in Figs. \ref{fig:5}(a)-\ref{fig:5}(i) is consistent with an intermediate/strong coupling regime, which means that the {weak coupling} model defined by Eq. (\ref{eq:H}) is not appropriate for describing the system. On the other hand, the fitting shown in Fig. \ref{fig:5}(j) suggests a system in the weak coupling regime, since $\gamma^\bullet\ll \Delta_0$, although the gap-closing feature is weak and barely visible in this case. Whether the system is strong or {weak coupling} obviously depends on all the materials, growth, and fabrication details of the SM-SC structures and cannot be decided \textit{a priori}.

\section{Conclusion}\label{sec:conclusion}

In this paper, we establish a direct relationship between the gap-closing feature, which characterizes the low-energy spectrum of a SM-SC hybrid structure in the topologically trivial regime, and key system parameters, such as the chemical potential and the spin-orbit coupling strength. Working within a pristine clean-wire (i.e., homogeneous system) assumption, we show that the curvature of the gap-closing feature is determined by the spin-orbit coupling and the chemical potential. In particular, we find that for a given value of the chemical potential there exists a ``critical'' spin-orbit coupling $ \alpha_c $ above which the gap-closing feature is entirely concave. Using both a weak coupling model and an intermediate/strong coupling self-energy approach, we find that this behavior is qualitatively the same in all SM-SC coupling regimes. Furthermore, based on our finding that the gap-closing feature has, in general, a nonzero curvature, we demonstrate that the effective $g$-factor is not directly related to the slope of this feature. In general, we caution against extracting an effective {$ g $}-factor from the gap closure features since such an average {$ g $}-factor is a property of the specific hybrid SM-SC structure (and depends crucially on the magnetic field and gate voltage regimes used in the experiment), and therefore, cannot tell us anything about how the Zeeman splitting varies in a different applied field regime, or in a different gate voltage regime or in a different SM-SC structure.  This is true even in clean nanowires without any extrinsic ABS complications.

Based on our analysis of the relationship between the curvature of the gap-closing feature and the system parameters, we propose a fitting procedure based on  simulated annealing  that allows one to extract effective parameters from experimentally measured low-energy conductance spectra.  To illustrate  the implementation of this scheme, we apply it to some recently reported experimental data.\cite{moor2018electric,vaitiekenas2018effective,grivnin2018concomitant} In particular, we find that, in general, the effective $g$-factor cannot be extracted directly from the slope of the gap-closing feature and we show that proximity-coupling to the parent superconductor results in a strong renormalization of the effective $g$-factor, which can be estimated quantitatively using fitting procedures based on different effective models. {We believe that the detailed analysis based on the minimal model presented in our paper has the potential for providing the effective nanowire parameters in realistic SM-SC hybrid systems through a careful fitting of the experimental data.  In particular, concavity or convexity of the gap-closing features as a function of the Zeeman field has important information regarding the effective $g$-factor and spin-orbit coupling strength of the semiconductor nanowire.
}

\begin{acknowledgments}
This work is supported by Laboratory for Physical Sciences and Microsoft. We also acknowledge the support of the University of Maryland supercomputing resources.
\end{acknowledgments}

\appendix
\setcounter{secnumdepth}{3}
\setcounter{equation}{0}
\setcounter{figure}{0}
\renewcommand{\theequation}{A\arabic{equation}}
\renewcommand{\thefigure}{A\arabic{figure}}
\renewcommand\figurename{Figure}
\renewcommand\tablename{Supplementary Table}

\section{The crossover spin-orbit coupling $\alpha_c$}\label{App_A}

In this appendix, we present the mathematical details of Eq.\eqref{eq:alphac}. In a semi-infinite-long nanowire($ x>0 $), the BdG Hamiltonian reads:
\begin{equation}\label{eq:H(k)}
 H(k)=(\eta k^2-\mu+k\alpha\sigma_z)\tau_z+V_Z\sigma_x+\Delta\tau_x, 
\end{equation}
where $ \eta=\frac{\hbar^2}{2m^*} $ and $ k=-i\frac{\partial}{\partial x} $. The other variables have the same definition as in Eq.\eqref{eq:H}. We seek for a general wavefunction which is the superposition of all the particular solutions, namely, $ \psi_n(x)=e^{-ik_nx}u_n $ with eigenenergy $ E $ satisfying
\begin{equation}\label{eq:Heigen}
\left[H(k_n)-E\right]u_n=0,
\end{equation}
where $ u_n $ is a four-dimensional Nambu spinor. 

The eight solutions of $ k_n $({$ n=1\dots8 $}) are symmetric over the real axis and imaginary axis given the in-gap energy due to the imposed symmetry in the Hamiltonian. \cite{huang2018metamorphosis} Since we are considering the bound states, the normalization condition $ \int_0^\infty dx \abs{\psi(x)}^2<\infty $ constrains that among the eight solutions to Eq.\eqref{eq:Heigen}, only the solution in the lower half of complex plane will be considered, namely $ k_1,-k_1^*,k_2,-k_2^* $, along with their eigenvectors $ u_1,u_2,u_3,u_4 $. The bound states also requires the boundary condition to be $ \psi(0)=\sum_{n=1}^{4}C_n u_n=0 $, which indicates the four corresponding eigenvectors $ u_1,u_2,u_3,u_4 $ are linear dependent. This is equivalent to define an ancillary function $ F(E) $,
\begin{equation}\label{eq:FE}
F(E)=\text{det}\left[u_1;u_2;u_3;u_4\right]=0.
\end{equation}

The four eigenvectors $ u_1,u_2,u_3,u_4 $ are essentially derived from theirs corresponding eigenvalues $ k_1,k_2,k_3,k_4 $, {which are just the solutions to Eq.\eqref{eq:Heigen}.} We may expand the characteristic equation explicitly {from Eq.\eqref{eq:Heigen}}, which is
\begin{widetext}
\begin{eqnarray}\label{eq:g}
&&\eta^4 k_n^8+\left[-2 \eta ^2 \left(\alpha ^2+2 \eta  \mu \right)\right] k_n^6+\left[\left(\alpha ^2+2 \eta  \mu \right)^2+2 \eta ^2 \left(\Delta ^2-E^2+\mu ^2-V_Z^2\right)\right] k_n^4\\
&+&\left[2 \alpha ^2 \left(\Delta ^2-E^2-\mu ^2+V_Z^2\right)+4 \eta  \mu  \left(-\Delta ^2+E^2-\mu ^2+V_Z^2\right)\right]k_n^2+E^4-2 E^2 \left(\Delta ^2+\mu ^2+V_Z^2\right)+\left(\Delta ^2+\mu ^2-V_Z^2\right)^2=0\nonumber
\end{eqnarray}
\end{widetext}

We take the four lower plane solutions and substitute them back to Eq.\eqref{eq:Heigen} to straightforwardly derive $ u_n $({$ n=1\dots4 $}), which will satisfy Eq.\eqref{eq:FE}. {Although $ F(E) $ is a function of all parameters, we are now only interested in the gap-closing features. Thus,} we emphasize its $ V_Z $-dependence by rewriting it to $ f(V_Z,E)=0 $. The implicit differentiation of $ f(V_Z,E) $ gives us the expression for effective $ g $-factor. To find the critical value of $ \alpha $, we seek for the situation where $ \max(g(V_Z))=0 $ in $ V_Z\in[0,V_{Zt}] $. {From Eq. \eqref{eq:g}, we impose a trial function $ \alpha^2=\chi\mu\eta $ and solve the coefficient $ \chi $ numerically. The result of $\chi  $ is actually the square of coefficient $ \beta $ up to a constant $ \frac{1}{2\sqrt{2}\pi} $ mentioned in Sec. \ref{sec:g}.}

\bibliography{curvature}

\end{document}